\documentclass{article}

\newcommand{\model}{\texttt{SonicVerse }}

\usepackage{aimc2025} 
\usepackage{array}
\usepackage{xcolor}
\usepackage{amsmath}

\usepackage[utf8]{inputenc} 
\usepackage[T1]{fontenc}    
\usepackage{hyperref}       
\hypersetup{
  colorlinks   = true, 
  urlcolor     = blue, 
  linkcolor    = black, 
  citecolor   = black 
}
\usepackage{url}            
\usepackage{booktabs}       
\usepackage{amsfonts}       
\usepackage{nicefrac}       
\usepackage{microtype}      
\usepackage{graphicx}       
\usepackage{authblk}

\title{SonicVerse: Multi-Task Learning for Music Feature-Informed Captioning}

\author[1]{Anuradha Chopra}
\author[1]{Abhinaba Roy}
\author[1]{Dorien Herremans}

\affil[1]{Singapore University of Technology and Design \par
\texttt{anuradha\_chopra@mymail.sutd.edu.sg, \{abhinaba\_roy, dorien\_herremans\}@sutd.edu.sg}}

\begin{document}

\maketitle

\begin{abstract}
Detailed captions that accurately reflect the characteristics of a music piece can enrich music databases and drive forward research in music AI. This paper introduces a multi-task music captioning model, \model \hspace{-.5em}, that integrates caption generation with auxiliary music feature detection tasks such as key detection, vocals detection, and more, so as to directly capture both low-level acoustic details as well as high-level musical attributes. The key contribution is a projection-based architecture that transforms audio input into language tokens, while simultaneously detecting music features through dedicated auxiliary heads. The outputs of these heads are also projected into language tokens, to enhance the captioning input. This framework not only produces rich, descriptive captions for short music fragments but also directly enables the generation of detailed time-informed descriptions for longer music pieces, by chaining the outputs using a large-language model. To train  the model, we extended the MusicBench dataset by annotating it with music features using MIRFLEX, a modular music feature extractor, resulting in paired audio, captions and music feature data. Experimental results show that incorporating  features in this way improves the quality and detail of the generated captions.

\end{abstract}







\section{Introduction}

Music captioning represents a challenging task that has been the focus of research efforts in recent years \citep{manco2021muscaps}. An automated system for accurate music captioning would be helpful in a myriad of ways. First, by allowing music feature capturing through a centralized system that outputs in organic formats, opening up the complex domain of music theory to a broader audience. Second, by potentially generating training datasets for music-to-text models, text-to-music models, and, music question-answering tasks \citep{liu2024music}. A system that produces detailed captions, describing both the musical characteristics of a piece and how they evolve over time, would allow the generation of datasets for text‑to‑music models, which can accept prompts that include the desired temporal evolution within pieces and convert them into generated compositions.

Ironically, this task remains an open challenge due to the lack of an extensive dataset. A number of approaches have attempted to mitigate this issue by augmenting smaller existing datasets with predefined modifications \citep{doh2023lp} \citep{melechovsky2023mustango}. Currently, the largest open-source dataset with music and matching text captions is JamendoMaxCaps \cite{roy2025jamendomaxcaps}, which contains over 362k songs. Other datasets are often much smaller, such as MusicBench containing around 50k audio samples \citep{melechovsky2023mustango}. Generative AI models typically need millions of data points, hence this motivates our study to create a captioning system that can generate captions with both music technical as well as more general musical feature descriptions for musical pieces. We propose that by training a model to generate detailed captions on short segments, these can then be chained using a Large Language Model to produce comprehensive captions that encompass the musical features as well as the temporal evolution of the piece.

Existing music captioning systems predominantly focus on generating captions that capture high-level, qualitative features such as the mood and potential setting suitable for the music. We refer to these as "soft features", in contrast to "hard features" such as key, instrumentation, vocals, or other technical musical components. Limited prior work has specifically targeted the extraction and inclusion of hard musical features in caption generation. Pseudo captioning techniques such as LP-MusicCaps \cite{doh2023lp} would only extract features and then use an LLM to rewrite them into a caption. Or to create the MusicBench dataset \cite{melechovsky2023mustango}, music features where first extracted and merged with the original caption using an LLM. They found that training the model on such an enriched dataset improved the model to include dedicated musical feature sin the captions. Directly integrating features into a captioning model  has only been done by \citet{gardner2023llark}, however, the accuracy of the music features extracted by some these systems has not been extensively evaluated. Open-source captioning systems lack the development of a comprehensive approach that generates captions incorporating detailed music theory components. Therefore, the creation of a system trained on creative commons data, capable of capturing concrete musical features and seamlessly integrating them into descriptive natural language outputs remains an open research challenge.

\begin{figure}[t]
    \centering
    \includegraphics[width=0.98\linewidth]{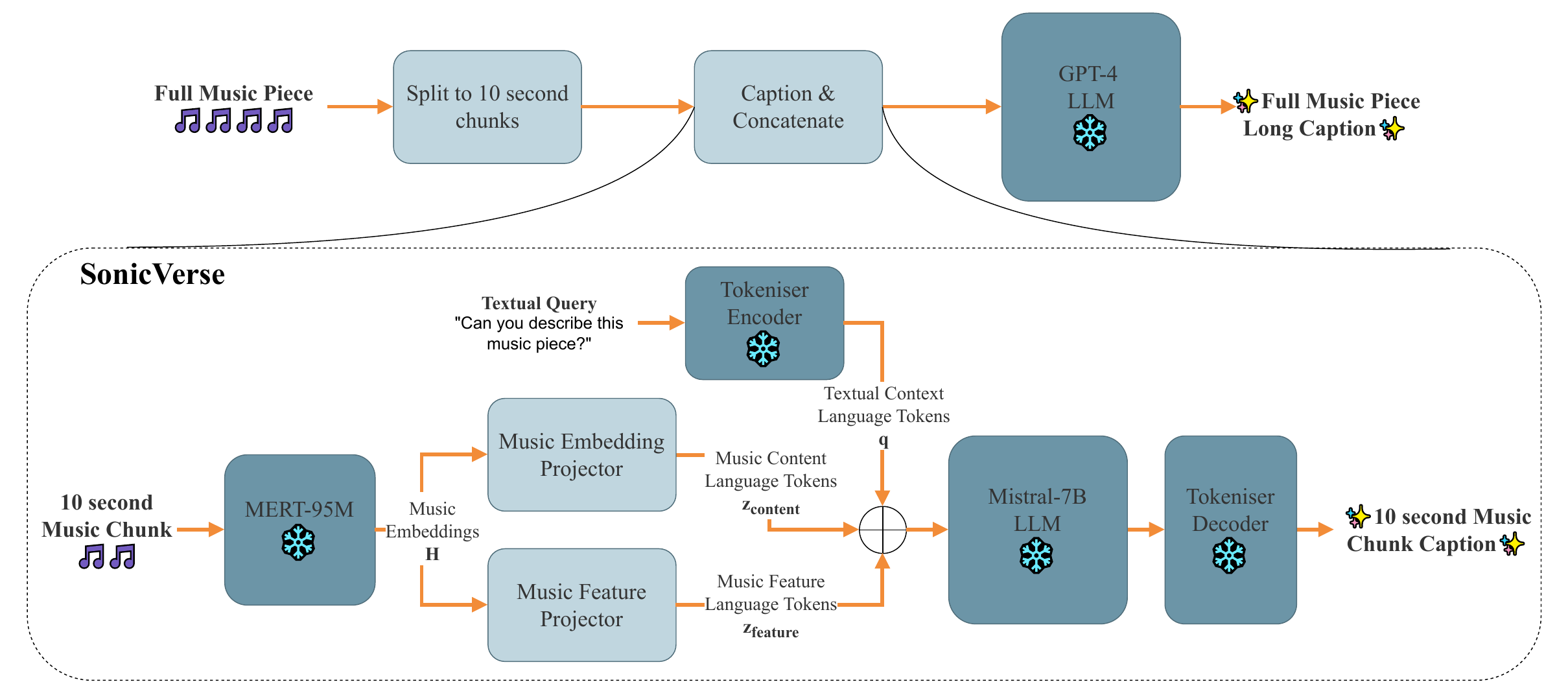}
    \caption{Overview of the proposed caption generation framework.}
    \label{fig:intro_pic}
\end{figure}

To address these questions, this paper proposes a novel caption generation model that is trained on a comprehensive dataset featuring labels for a wide range of concrete musical features, such as key, instrumentation, genre, mood, vocals, beats, and chords. The model employs a multi-task projector that operates on the music embeddings to output feature extractor-guided tokens for input into a large language model for caption generation. By utilizing a multi-task learning algorithm with auxiliary heads to predict the individual feature-related information, the model is able to convert these extracted features into language tokens that can be effectively incorporated into the captioning process.

Furthermore, the music embeddings are also directly converted into language tokens, enabling the capture of more abstract, qualitative "soft" features to enhance the descriptiveness of the generated captions. Experiments conducted in this study demonstrate that the incorporation of these music feature extractors within the token projection model leads to significant improvements in the quality of the generated captions, as measured by n-gram metrics such as BLEU \citep{papineni2002bleu}, ROUGE \citep{lin2004rouge}, and METEOR \citep{banerjee2005meteor}. Finally, We explore how we can generate temporally-informed long captions for full-length pieces of music by leveraging LLM-chaining.

The primary contributions of the paper are as follows:

\begin{enumerate}
    \item  We propose a novel multi-task learning framework that jointly learns music captioning and musical feature prediction. This approach enables effective training on smaller, open-source datasets by leveraging auxiliary supervision.
    
    \item  We introduce a chaining mechanism with large language models (LLMs) to generate temporally-aware, coherent captions for long-form music. This allows our model to describe the musical progression over time using fine-grained clip-level captions.
     
    \item Our architecture eliminates the need for external music feature extractors by integrating feature prediction directly into the captioning pipeline. The full model, \model \hspace{-.5em}, is released as open source to support reproducibility and future research\footnote{The model with online demo is available at \url{https://github.com/AMAAI-Lab/SonicVerse}}. 
\end{enumerate}

The rest of the paper is structured as follows. Section 2 discusses the related work. Section 3 details the method and architecture of our model. Followed by section 4 that describes the experimental setup, including the dataset and metrics used. Finally, Section 5 discusses the results followed by conclusion.

\section{Related Work}

Music captioning techniques can be viewed as a sub-domain of the broader field of audio captioning. Consequently, many methods developed for music captioning draw from general audio description approaches. Existing research can broadly be classified into three categories: pseudo-caption generation using music tags and large language models (LLMs), encoder-decoder frameworks for cross-modal translation, and token projection methods that integrate music representations into LLMs.

One class of methods focuses on generating pseudo-captions from music feature labels and tags, which are reformulated into structured prompts for LLMs to generate final captions \citep{doh2023lp}. These approaches do not process raw audio themselves and instead rely entirely on dedicated feature extractors. A recent example is LP-MusicCaps \citep{doh2023lp}, which addresses the lack of annotated data by using GPT-3.5 Turbo to create a synthetic dataset of over 2.2 million pseudo-captions for over 500k audio clips. This dataset comprises subsets such as LP-MusicCaps-MC, built from 10-second clips in the MusicCaps dataset \citep{agostinelli2023musiclm}; LP-MusicCaps-MTT, using 30-second clips from Magna-Tag-A-Tune \citep{law2009evaluation}; and LP-MusicCaps-MSD, drawing on the Million Song Dataset \citep{dataset2011million}. Captions are generated from features through prompts for the four predefined tasks: writing, summarisation, paraphrasing, and attribute prediction. They are evaluated using both automatic metrics (BLEU \citep{papineni2002bleu}, METEOR \citep{banerjee2005meteor}, ROUGE-L \citep{lin2004rouge}, BERT-Score \citep{zhang2019bertscore}) and human listening tests. A cross-modal model trained on LP-MusicCaps has demonstrated strong zero-shot and transfer learning performance \citep{doh2023lp}. However, the quality of the generated captions is fundamentally bounded by the prompt design, the accuracy of the music feature prediction models, and the inherent capabilities of the LLM, since these systems lack direct audio input and therefore lack any direct acoustic cues. 

A second group of methods uses encoder-decoder frameworks to learn mappings between music and text representations through latent spaces. For example, the Synaesthesia model \citep{kuang2022music} employs a dataset of 1,955 Western classical recordings paired with professional descriptions to train a cross-modal translation model. Audio and text are encoded separately and aligned using a group topology preservation loss \citep{kuang2022music}, encouraging semantic consistency. While the method benefits from aligned professional annotations, its reliance on classical music limits generalization to other genres. Furthermore, the authors point out that the generated captions occasionally exhibit formulaic patterns and grammatical inconsistencies.

A third and increasingly prominent category involves projecting encoded music features into the token space of LLMs, allowing for joint inference over both audio and text inputs. MusiLingo \citep{deng2023musilingo}, for instance, uses the MERT-330M encoder \citep{li2023mert} to extract music representations, which are then projected into token embeddings for a frozen Vicuna-7B model. Pretraining is performed on the LP-MusicCaps-MSD dataset \citep{doh2023lp} for captioning, followed by fine-tuning on music question-answering using the MusicInstruct dataset \citep{deng2023musilingo}. Similarly, the Music Understanding LLaMA (MU-LLaMA) model \citep{liu2024music} maps MERT-encoded audio \citep{li2023mert} through a custom adapter into the input space of Meta’s LLaMA model. The model is used for augmenting captioning datasets such as MusicCaps \citep{doh2023lp} and MagnaTagATune \citep{law2009evaluation} with synthetic Q\&A pairs generated by MosaicML MPT-7B \citep{MosaicML2023Introducing}. The augmented dataset is used to train a multi modal interface for captioning and reasoning over music.

Several other models build on similar principles. For instance, SALMONN \citep{tang2023salmonn} combines Whisper \citep{radford2023robust} and BEATs \citep{chen2022beats} encoders with a window-level Q-former to extend LLMs for general audio understanding, including speech and music tasks. The LLaRK model \citep{gardner2023llark}, on the other hand, incorporates explicit music attributes like key, tempo, and instrumentation. It uses Jukebox-5B \citep{kamuni2024exploring} as the encoder and a multimodal projection layer for the incorporation of music features along with the music encodings. However, LLaRK’s weights are not publicly released. Other notable work includes BLAP (Bootstrapping Language–Audio Pre-training) \citep{lanzendorfer2025bootstrapping}, which aims to match the performance of large-scale captioning systems using significantly less data. This model combines the CLAP encoder \citep{laionclap2023} with a frozen Flan-T5 model and a Q-Former for latent space alignment. While it can achieve good comparative results with lesser data and a smaller architecture, it tends to produce concise captions with lesser details, as compared to the reference MusicCaps.
FUTGA \citep{wu2024futga} introduces temporally-enhanced generative augmentation to generate fine-grained, segment-level captions by aligning synthetic labels with human annotations from HarmonixSet \citep{nieto2019harmonix}. This is achieved by automatically annotating smaller segments of two- to five-minute audio clips and aligning synthetic captions with human annotations from HarmonixSet \citep{nieto2019harmonix}, FUTGA enables time-localized descriptions that aid retrieval and segment-aware generation tasks. While capable of identifying structural segments of music such as intro, verse, chorus, etc. the weights for BEATs \citep{chen2022beats} are unavailable publicly at the time of writing. Finally, the recent QWEN2-Audio model by Huawei \citep{chu2024qwen2}, is a large scale audio language model (LALM), which was trained on a large amount of speech (370k hours), sound (10k hours), and music (140k hours) data. The model uses an encoder based on  Whisper-large-v2 and frozen Qwen-7B LLM, and is evaluated on the AIR-Bench Chat-Benchmark-Music dataset \citep{yang2024air}. This large model outperforms other LALMs such as SALMONN \citep{tang2023salmonn}. We note that the evaluation process does not capture accuracy of the music features detected, but rather measures for overall descriptiveness and coherence.

Although MusiLingo \citep{deng2023musilingo}, MU‑LLaMA \citep{liu2024music}, and BLAP \citep{lanzendorfer2025bootstrapping} provide publicly available weights, none offer an explicit music-feature extraction module. SALMONN \citep{tang2023salmonn} and FUTGA \citep{wu2024futga} depend on the BEATs encoder \citep{chen2022beats}, whose weights are not currently available on their hosted links. LLaRK \citep{gardner2023llark} focuses on structured music attributes but does not release its projection layers, and QWEN2‑Audio \citep{chu2024qwen2} relies on large proprietary datasets and unpublished code. In addition, many of these models, with the exception of FUTGA \cite{wu2024futga}, do not generate time-informed long form captions for full length audio. 

\begin{figure}[t]
    \centering
    \includegraphics[width=0.98\linewidth]{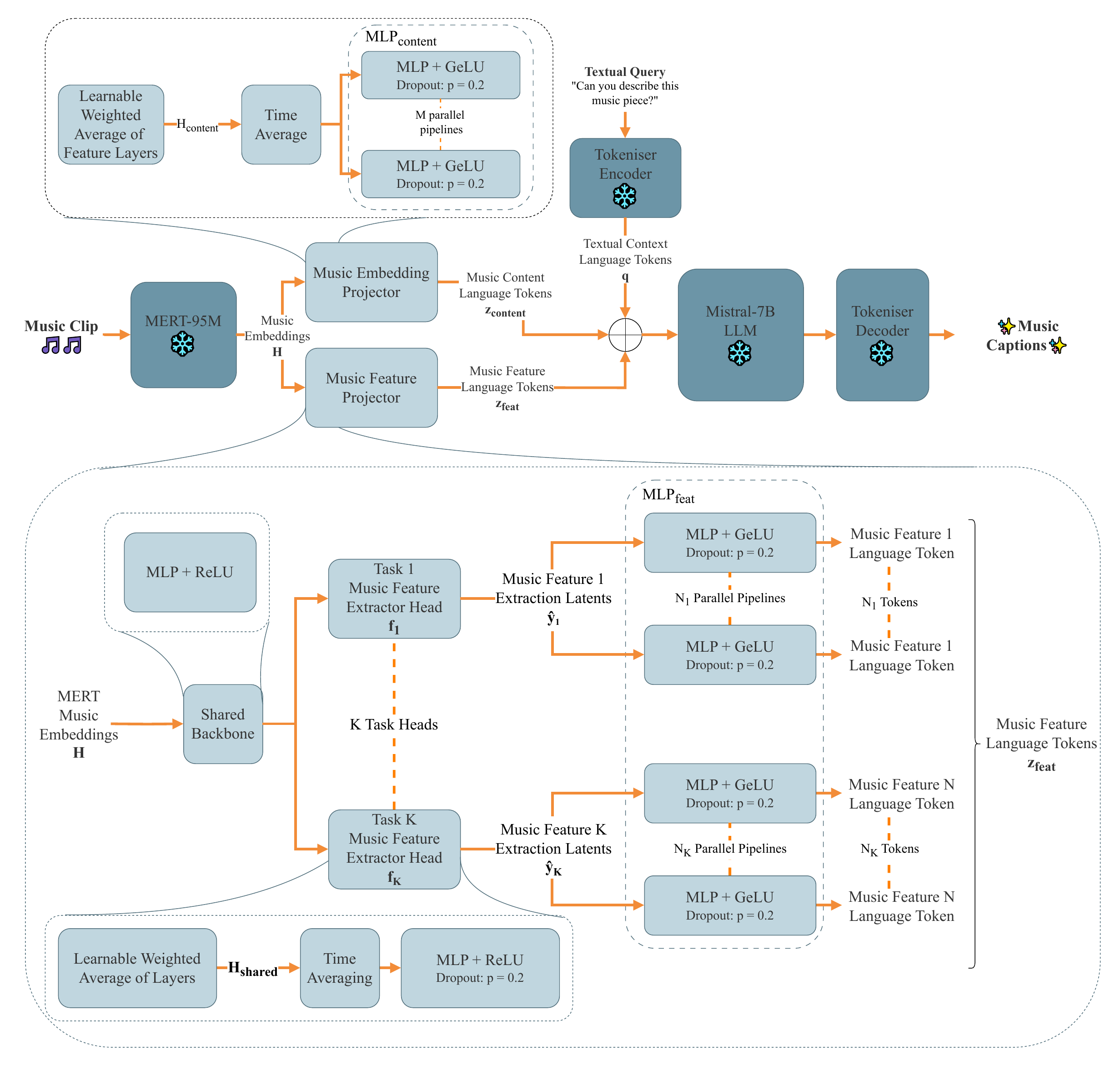}
    \caption{Overview of the proposed SonicVerse architecture.}
    \label{fig:main_figure}
\end{figure}

In this work, we aim to address these gaps by releasing both our model and its weights. Our architecture (see Figure!\ref{fig:main_figure}) explicitly targets the detection and integration of detailed musical features and is trained exclusively on creative commons datasets, ensuring full reproducibility as well as legality. Moreover, it also provides a mechanism to caption short chunks of a long music piece, and chain it together, to retrieve long captions that capture the temporal evolution of the input audio with general description as well as music features.

\section{Proposed music captioning model}
In this section, we outline the details of the proposed \model \hspace{-.5em} architecture. Our aim is to capture high-level music features, such as instrumentation and key, during the captioning task. To achieve this, we employ a multi-task learning approach that integrates the latent representations of these concrete music features directly into the captioning pipeline.

The proposed model learns to predict various musical features with a multitask setup, and projects them as input for the language model. Concurrently, the original music encodings are also projected to form input language model tokens. These inputs are concatenated and the resulting dual representation provides latent representations of the music clip as a whole for learning general features, while also incorporating the specific music features to guide the captioning process. Finally, these projected tokens are concatenated with the query tokens and passed into a pre-trained large language model to generate the final caption.

Mathematically, let \( x \in \mathbb{R}^{T \times F} \) denote the input music clip, where \( T \) is the number of audio frames and \( F \) the feature dimension. Our model processes this input and produces a caption using:

\begin{equation}
\texttt{Caption} = \texttt{LLM} \left( \left[ \mathbf{z}_{\text{content}} \, \| \, \mathbf{z}_{\text{feature}} \, \| \, \mathbf{q} \right] \right)
\end{equation}

where \( \mathbf{z}_{\text{content}} \in \mathbb{R}^{M \times d} \) and \( \mathbf{z}_{\text{feature}} \in \mathbb{R}^{N \times d} \) are language tokens derived from music content and music features respectively, \( \mathbf{q} \) represents textual query tokens, and \( d \) is the embedding dimension of the pre-trained language model. The variable \(M\) is the number of tokens representing the music content and \(N\) the number of tokens representing the textual query.

The overall architecture of our music captioning model consists of three primary components: 1) Music Encoder, 2) Multi-tasking Projector, and 3) Pre-trained Large Language Model.
We describe the design choices and justification for each of these elements below.

\subsection{Music Encoder}

To select the encoder, we first consider the ability to extract information required for language token prediction, from audio clips, as well as capturing musical attributes such as instruments, genre, mood/theme, key, vocals/instrumental, and vocals gender. Second, we looked at the ability to preserve feature information while still producing efficient embeddings.  

As the audio encoder, we chose the use MERT (Music undERstanding model with large-scale self-supervised Training) \citep{li2023mert}. MERT leverages multi-task learning, using RVQ-VAE for acoustic information and CQT for musical information. It consists of a transformer architecture with self-attention, to model long range dependencies and hierarchical music patterns. MERT has been validated on 14 diverse MIR (music information retrieval) tasks and achieved state-of-the-art performance. 
The MERT encoder outputs the hierarchical embedding $\mathbf{H}$:

\begin{equation}
\mathbf{H} = \texttt{MERT}(x) \in \mathbb{R}^{L \times T' \times D}
\label{eq:mert}
\end{equation}

where \( L = 13 \) is the number of representation layers, \( T' \) is the temporal resolution, and \( D = 768 \) is the embedding dimension.
Each layer captures dedicated acoustic information, which could be related to musical features such as tonal key or genre. For instance, the authors of MERT report that the higher layers do not perform particularly well for the genre detection task \citep{li2023mert} and suggest empirically choosing layers for tasks or using a learnable weighted average\footnote{\url{https://huggingface.co/m-a-p/MERT-v0}}.

\subsection{Multi-task projector}

The multi-layered output of the MERT encoder, and its ability to capture diverse musical attributes make it a natural choice for our music captioning architecture. The projector component of our model plays a crucial role in leveraging these rich music representations to inform the captioning process. 
The module passes each fixed length audio clip through the MERT encoder in Equation~\ref{eq:mert} to obtain the embeddings $\mathbf{H}$. These MERT embeddings are then fed into a multi-task projection architecture consisting of a music content projector (${H}_{\text{content}}$) and a music feature projector (${H}_{\text{feature}}$) pathway as detailed in Figure~\ref{fig:main_figure}. 

First, the \textbf{music embedding projector } computes a learned layer-weighted average combined with a time-averaged representation of the MERT features and passes it through parallel fully connected layers and GeLU units. The approach to use MERT embeddings as input to the feature projector using a learned layer-weighted average was selected based on the suggestions of the MERT authors\footnote{\url{https://huggingface.co/m-a-p/MERT-v0}}. The shape of the MERT embeddings is initially \(  L \times T' \times D \). After time averaging, the shape of the embeddings becomes \(  L  \times D \), thus eliminating the time dimension, resulting in a fixed number of tokens regardless of the input audio length.

The learned weighted-averaged time-averaged embeddings are passed through a set of parallel  multi-layer perceptrons with GeLU activations, \( \texttt{MLP}_{\text{content}} \). The result is a set of tokens, \(z_{content}\), suitable as input for the frozen LLM.


\begin{equation}
\mathbf{H}_{\text{content}} = \sum_{\ell=1}^{L} \alpha_{\ell} \cdot \mathbf{H}[\ell], \quad \text{with} \quad \sum_{\ell=1}^{L} \alpha_{\ell} = 1, \quad \alpha_{\ell} \geq 0
\end{equation}

\begin{equation}
\mathbf{z}_{\text{content}} = \texttt{MLP}_{\text{content}} \left( \frac{1}{T'} \sum_{t=1}^{T'} \mathbf{H}_{\text{content}}[t] \right)
\end{equation}

Here, \(\alpha_{\ell}\) represents the learned weights for the representation layer \(\ell\) which is \(\leq L\), while \(H[\ell]\) represents the \(\ell^{th}\) layer of \(H\).
The tokens, \(z_{content} \), while not representative of textual word embeddings directly, are learned against the frozen LLM's embedding by training on known captions. This ensures that they are learned such that they are aligned with English words that describe the music content.

Secondly, the \textbf{music feature projector} is a multi-task network that captures discrete musical attributes via a shared backbone. For this shared backbone, we also compute the learned layer-weighted average combined with a time-averaged representation of the MERT features, represented by \(H_{shared}\). 

\begin{equation}
\mathbf{H}_{\text{shared}} = \sum_{\ell=1}^{L} \beta_{\ell} \cdot \mathbf{H}[\ell], \quad \text{with} \quad \sum_{\ell=1}^{L} \beta_{\ell} = 1
\label{eq:h_shared}
\end{equation}

In Equation~\ref{eq:h_shared}, \(\beta_{\ell}\) represents the learned weights for the representation layer \(\ell\) which is \(\leq L\), while \(H[\ell]\) represents the \(\ell^{th}\) layer of \(H\).
Let the total number of music features that are predicted in the music feature extractor be \(K\).
Each task head \( f_k \) (for \( k  = 1, \dots, K \)) predicts a set of probabilities, \(\hat{\mathbf{y}}_k\) for a specific music feature:

\begin{equation}
\hat{\mathbf{y}}_k = \sigma \left( f_k \left( \frac{1}{T'} \sum_{t=1}^{T'} \mathbf{H}_{\text{shared}}[t] \right) \right), \quad \hat{\mathbf{y}}_k \in [0, 1]^{C_k}
\end{equation}

Next, each music feature vector is passed through a set of parallel multi-layer perceptrons with GeLU activations, represented by \( \texttt{MLP}_{\text{feat}} \). The output of each multi-layer perceptron is an individual input token for the frozen LLM. All output tokens are concatenated to form a set of tokens \(z_{\text{feature}}\), that represent the predicted music features, projected into the language space. These output tokens are learnt against the frozen LLM's embedding, and become aligned with English words that describe the music features, as they are trained with known English captions.

\begin{equation}
\mathbf{z}_{\text{feature}} = \texttt{concat} \left( \texttt{MLP}_{\text{feat}}\left( \hat{\mathbf{y}}_k\right) \right), \quad k \in [1, K]
\end{equation}

The latent vectors $\mathbf{z}_{\text{content}}$ and $\mathbf{z}_{\text{feature}}$ are then combined with with the textual query tokens $\mathbf{q}$ and fed into the pre-trained LLM (Mistral-7) \citep{jiang2023mistral} to generate the captions.

Finally, the \textbf{training objective} is defined as the weighted sum of the captioning and feature classification losses:

\begin{equation}
\mathcal{L} = \lambda_{\text{cap}} \cdot \mathcal{L}_{\text{cap}} + \sum_{k=1}^{K} \lambda_k \cdot \mathcal{L}_k
\end{equation}

where:
\begin{itemize}
    \item \( \mathcal{L}_{\text{cap}} \) is the cross-entropy loss for captioning.
    \item \( \mathcal{L}_k = \texttt{BCEWithLogits}(\hat{\mathbf{y}}_k, \mathbf{y}_k) \) is the binary classification loss for task \( k \).
    \item \( \lambda_{\text{cap}}, \lambda_k \) are task weighting hyperparameters.
\end{itemize}

The key innovation in this projector architecture is the incorporation of music feature extractor heads within the music-to-language token projection, which directly infuses concrete musical attributes into the token representations to guide the captioning process.

\subsection{LLM-enabled chaining}

The above architecture generates captions for short fragments of music (10s or 30s segments in our training dataset). We further leverage these captions to generate elaborate captions that describe the temporal evolution throughout a long piece of music, as shown in Figure~\ref{fig:intro_pic}. To achieve this, the full-length song is first clipped into fixed length (10 second) chunks, which are input into the model described above. After retrieving captions for all of the chunks, they are combined with a curated prompt that requests the generation of a detailed description of the entire song, while retaining musical feature information. The prompt is fed into a state-of-the-art LLM. In our experiments, we opted to use GPT-4 \citep{achiam2023gpt} for the generation of a single long caption that captures the music feature details and their evolution over the length of the music piece.

Mathematically, each audio segment $x$ is divided into \( N \) fixed length clips \( \{x^{(i)}\}_{i=1}^N \), generating chunk level caption \(c^{(i)}\).
\begin{equation}
    c^{(i)} = \texttt{\model \hspace{-.5em}}(x^{(i)})
\end{equation}

The chunk level captions are concatenated into a structured prompt \( p \) to give the long caption.

\begin{equation}
    \texttt{LongCaption} = \texttt{GPT4}(p)
\end{equation}

The prompt used to generate the caption through chaining is given below. 

\begin{verbatim}
    Given the following chronological 10‑second chunk descriptions of a 
    single piece, write one flowing, detailed description of the entire song
    —its structure, instrumentation, and standout moments. Mention transition 
    points in terms of time stamps. If the description of certain chunks does 
    not seem to fit with those for the chunks before and after, treat those 
    as bad descriptions with lower accuracy and do not incorporate the information. 
    Retain concrete musical attributes such as key, chords, tempo.

    Chunks for “{song_name}" :
    1. 0 to 10 seconds: {Chunk 1 caption}
    2. 10 to 20 seconds: {Chunk 2 caption}
    ...
    Full song description:
\end{verbatim}

\section{Experimental setup}

We discuss the dataset used, baselines, and implementation details in this section.

\subsection{Dataset}

We use three datasets for training the proposed architecture. First, we use the Jamendo dataset \citep{bogdanov2019mtg}, which consists of a collection of around 55k music clips (30s) with tags for instrument, mood/theme and genre. We extend this dataset to include labels for vocals, vocals gender, and key, using MIRFLEX \citep{chopra2024mirflex}. Secondly, for general music-language alignment, we train the model using the Magna-Tag-A-Tune dataset \citep{law2009evaluation}, which consists of around 25k music clips (30s), with paired captions generated using MosiacML \citep{MosaicML2023Introducing} for the training of the captioning MuLLaMA \citep{liu2024music}. We extend this dataset using MIRFLEX \citep{chopra2024mirflex} to get labels for instrument, mood / theme, genre, key, vocals and vocals gender. Finally, we use the MusicBench training set which consists of  around 26k clips (after excluding those samples that are part of MusicCaps eval set) with  paired captions (10s). These captions explicitly include music features. This dataset was also extended using MIRFLEX to obtain labels for instrument, mood / theme, genre, key, vocals and vocals gender.

\subsection{Baselines}
\label{sec:baselines}
To assess the impact of incorporating explicit music feature representations within our music-to-token projector, we performed an ablation study, comparing two baselines: 

\textbf{Baseline A}: 'Control' baseline uses only the music context projector,

\textbf{Baseline B} : 'Music feature projector augmented' baseline extends the control baseline by integrating the music feature extraction heads followed by the music feature to language token projection, thus creating our proposed \model \hspace{-.5em}.

The section below describes the training procedures and evaluation metrics used for both baselines.

\subsection{Implementation details}

We fixed the number of tokens \(N + M\) to $60$ across all experiments. In the control baseline (Baseline A), a single MLP projector is used to map aggregated music embeddings directly to language tokens. In the music feature projector augmented baseline (Baseline B) we allocated \(M = 35\) tokens for the music embedding projection and \( N_k = 5 \) tokens for each task head, namely key detection, instrument detection, mood / theme detection, genre detection, vocals and vocals gender detection.

Both baselines are pretrained on the Magna-Tag-a-Tune \citep{law2009evaluation} + MosiacML datasets \citep{MosaicML2023Introducing}, and then finetuned on the MusicBench Training Set \citep{melechovsky2023mustango}. Because the datasets lack ground-truth music feature labels, we augment them using MIRFLEX \citep{chopra2024mirflex}, which generated labels for key, instrument, mood/theme, genre, vocals and vocals gender, chords as well as downbeat timings. This process provides us with a dataset of aligned triplets of audio, text caption, and feature labels for the multi-task training of Baseline B. We note that using the MIRFLEX feature extractors rather than ground truth music features might introduce some noise and bias into the dataset. However, the final evaluation of the model is done on the ground trust \textit{caption}, which is not dependent on the extracted features. When we evaluate the accuracy of the presence of music features in the generated captions in Section~\ref{tab:music_metrics_comparison_with_prior_work}, we compare to these to the original captions without referring to MIRFLEX.

Specific to Baseline B, the feature extractor is first pretrained with the Jamendo dataset \citep{bogdanov2019mtg} extended using MIRFLEX \citep{chopra2024mirflex} and finetuned on augmented music features in the Magna-tag-atune \citep{law2009evaluation} + MosiacML dataset. A weighted sum of the losses for each task is used as the back-propagated loss, with weight \( \lambda_k = 0.2 \)for each task (instrument detection, mood detection, genre detection, key detection, vocals detection). 

During the captioning pretraining step, we employ the extended Magna-Tag-aTune dataset which is comprised of 30s music clips. The loss is weighted with \( \lambda_{cap} = 1.0 \) for the captioning loss and \( \lambda_k = 0.1 \) for all individual feature extraction task individually. The captioning model is then finetuned on the extended MusicBench dataset. 
To evaluate the performance of both configurations, we used the traditional metrics from natural language processing (NLP) as well as designed some music-specific metrics.

\subsection{Evaluation metrics}

All text caption (NLP) evaluations were performed on the MusicCaps evaluation set \citep{agostinelli2023musiclm}, while all music-feature specific evaluations were performed on the MusicBench test set \citep{melechovsky2023mustango}, following the MusicCaps split \citep{agostinelli2023musiclm}. 

\paragraph{NLP Metrics} Similar to other captioning papers \citep{doh2023lp, kuang2022music, deng2023musilingo, liu2024music, tang2023salmonn}, we used metrics that are generally used for assessing translation accuracy and sentence similarity. To assess the accuracy of the generated captions against ground-truth annotations  we used the following key metrics: 

\begin{itemize}
    \item \textbf{BLEU Score} \citep{papineni2002bleu} quantifies the overlap between the predicted and reference captions by matching $n$-grams (up to a certain length).
    \item \textbf{BLEU-4 Score} \citep{papineni2002bleu} focuses on a 4-gram comparison making it more suitable for longer, complex sentences and phrases.
    \item \textbf{METEOR Score} \citep{banerjee2005meteor} aligns predictions with references, using synonyms, stemming and exact matching. It penalises longer captions that do not match significantly.
    \item \textbf{ROUGE Score} \citep{lin2004rouge} measures the overlap of n-grams, word sequences and word pairs, prioritising recall over precision score. It measures how much of the reference appears in the prediction.
    \item \textbf{BERT Score} \citep{zhang2019bertscore} measures semantic similarity through the use of BERT embeddings. It captures the context and meaning rather than matching words.
\end{itemize}

\paragraph{Music Metrics} While the NLP metrics may give us a measure to compare similarity of captions, they do not explicitly capture how much of the music features are accurately mentioned in the caption. To enable this evaluation, we designed metrics that provide insight in the overlap between the (correct) mention of the key, instrumentation, and vocals between the caption and audio. These metrics are obtained using a state-of-the-art LLM (GPT-4) with the below prompt that requests a comparison of the reported features across the prediction and reference.

\begin{verbatim}
You are tasked with comparing two descriptions of a musical piece. 
Evaluate the following aspects:

1. Key Match: Does the musical key specified in the prediction match that 
in the reference? If missing in reference, respond 'n/a'.
2. Instrumentation Match: Do the instruments described in the prediction 
correspond to those in the reference? If missing in reference, respond 'n/a'.
3. Genre Match: Does the genre implied by the description in the prediction 
match that in the reference? If missing in reference, respond 'n/a'.
4. Mood/Theme Match: Does the mood or theme of the music in the prediction 
match that in the reference? If missing in reference, respond 'n/a'.
5. Vocal Presence Match: Does the presence or absence of vocals in the 
prediction match that in the reference? 
6. Vocal Gender Match: If vocals are present, does the gender of the vocals 
(male or female) in the prediction match that in the reference? If missing 
in reference, respond 'n/a'

Return your answer as a JSON object with the following keys:
    'key_match', 'instrument_match', 'genre_match', 
    'mood_match', 'vocal_presence_match', 'vocal_gender_match'.

Values should be 'yes', 'no', or 'n/a' if the attribute is not applicable.

---
Prediction:
{prediction_text}

Reference:
{reference_text}
\end{verbatim}

The resulting accuracy for each feature is then calculated as the total number of matches (output yes) divided by the total number of valid pairs (output yes or no, excluding all n/a output).

\section{Results}

In this section, we report results of our ablation study, as well as compare the performance of our model with state-of-the-art models, and analyse an example output of a caption for a full-length song.

\subsection{Ablation study}

In order to show the importance of guiding the caption with the use of auxiliary music feature detection tasks, we conduct an ablation study using the baselines described in Section \ref{sec:baselines}.
The results are shown in the Table \ref{tab:ablation}. The inclusion of music feature–projected tokens in Baseline B leads to superior performance across all evaluated NLP metrics except METEOR \citep{banerjee2005meteor}, suggesting that our feature‑augmented projector better captures musical nuances in the generated captions. 

\begin{table}[h!] 
    \caption{Ablation study results on the MusicBench test set (MusicCaps split)}
    \label{tab:ablation}
    \centering    \begin{tabular}{lccccc}
    \toprule
          \textbf{Model}&\textbf{BLEU}$\uparrow$&  \textbf{BLEU-4 }$\uparrow$&\textbf{METEOR}$\uparrow$&\textbf{ROUGE}$\uparrow$&\textbf{BERT}$\uparrow$\\
          \midrule
 Baseline A & 0.3456& 0.1799& \textbf{0.2507}& 0.2621&0.8716\\
 Baseline B (\model \hspace{-.5em}) & \textbf{0.3484}& \textbf{0.1824}& 0.2506& \textbf{0.2622}&\textbf{0.8723}\\
\bottomrule
\end{tabular}
\end{table}

\subsection{Comparison with state-of-the-art}

We present a comparison with state-of-the-art models in Table~\ref{tab:music_captioning_comparison_with_prior_work} in terms of NLP metrics. All of the models are evaluated on the MusicCaps Test dataset. The results were taken from the respective papers, except BLAP and QWEN2-Audio. The latter two models were run to obtain the values. When examining the table, we should keep in mind that models such as QWEN2-Audio were trained on large private datasets (potentially including our test set MusicCaps). The table thus serves as a rough benchmark. It is good to observe that \model \hspace{-.5em} can outperform most models, even recent ones such as BLAP that are also trained on limited datasets. This confirms that our proposed model is able to generate high-quality music captions. 

\begin{table}[h!]
    \centering     \caption{Results of state-of-the-art music captioning models on the MusicCaps test set.}
    \label{tab:music_captioning_comparison_with_prior_work}
    \begin{tabular}{lccccc}
    \toprule
          \textbf{Model}&\textbf{BLEU$\uparrow$}&  \textbf{BLEU-4$\uparrow$}&\textbf{METEOR$\uparrow$}&\textbf{ROUGE$\uparrow$}&\textbf{BERT$\uparrow$}\\
          \midrule
          LP-MusicCaps &-& 
     0.0605& 0.2239&0.1303&0.8451\\
 MusiLingo & 0.308& -& 0.216& 0.217&0.868\\
 SALMONN & -& 0.055& -& 0.218&-\\
 LLARK & 0.28& 0.14& 0.28& 0.25&-\\
 QWEN2-Audio
 & 0.3427& \textbf{0.2109}& \textbf{0.2584}& 0.2405&\textbf{0.8816}\\
BLAP
 & 0.1024& 0.0406& 0.0746& 0.0688&0.8502\\
 \model & \textbf{0.3484}& 0.1824& 0.2506& \textbf{0.2622}&0.8723\\
\bottomrule
\end{tabular}
\end{table}

The metrics above, however, capture only the NLP similarities. As discussed in the previous section, we have designed dedicated metrics to assess if the caption accurately captures specific musical aspects such as key, instrument, genre, mood, vocals and gender of vocals. Table~\ref{tab:music_metrics_comparison_with_prior_work} shows the result of these metrics on the MusicBench test set for QWEN2-Audio, BLAP, as well as \model \hspace{-.5em}. We chose to compare with QWEN2-Audio as it is a large model representing the state-of-the-art trained on private data. BLAP, on the other hand, represents a smaller, open model trained on open data. The MusicBench test set was selected here, as annotations for the music features are available. In terms of captioning the correct key, \model \hspace{-.5em} outperforms all other models, however, as expected the model trained on `internal' data (QWEN2-Audio) takes the lead when it comes to most other features. The proposed \model \hspace{-.5em} does show competitive performance when comparing to the other model trained on open data (BLAP) as it outperforms on all but one of the music metrics. This again confirms the ability of our model to successfully capture music features in the captions, facilitated through the multi-task mechanism. 

\begin{table}[h!]
    \centering  
    \caption{Results of the music metrics on the MusicBench test set (MusicCaps split).}\label{tab:music_metrics_comparison_with_prior_work}
    \begin{tabular}{>{\raggedright\arraybackslash}p{0.16\linewidth}>{\raggedright\arraybackslash}p{0.1\linewidth}>{\centering\arraybackslash}p{0.11\linewidth}>{\centering\arraybackslash}p{0.1\linewidth}>{\raggedright\arraybackslash}p{0.1\linewidth}>{\raggedright\arraybackslash}p{0.1\linewidth}>{\raggedright\arraybackslash}p{0.1\linewidth}}
    \toprule
          \textbf{Model}&\textbf{Key$\uparrow$}&  \textbf{Instrument$\uparrow$}&\textbf{Genre$\uparrow$}&\textbf{Mood$\uparrow$}&\textbf{Vocals$\uparrow$}&\textbf{Gender$\uparrow$}\\
          \midrule
          BLAP & 0.2162 & 0.3231& 0.2987& 0.4643&0.7656 & \textbf{0.7797}\\
           QWEN2-Audio & 0.2966 & \textbf{0.6582}& \textbf{0.5486}& \textbf{0.6389}&\textbf{0.8808}& 0.7654\\
           \model & \textbf{0.3048}& 0.4337& 0.4329& 0.4725&0.8051 & 0.7764\\
\bottomrule
\end{tabular}

\end{table}

\subsection{Case study -- LLM-enabled caption chaining}

To demonstrate the capability of our proposed \model \hspace{-.5em} framework to generate long form captions, we  examine a case study. We use the popular music piece `Bohemian rhapsody' by Queen and feed it through our pipeline. The resulting caption is given below, and clearly captures the temporal evaluation through the piece. By training our model to recognise detailed musical features of short 10s fragments, this example demonstrates that LLM chaining can then be used to expand those detailed captions into a temporally informed caption. For instance, the first segments mentions the `choir' part of Bohemian Rhapsody, whereas after 80s the song `morphs into a love song'. This is a very apt description of the feeling of those segments and how they change over time. 
In addition, our framework offers flexibility as the resulting output caption can easily be customized tweaked to focus on certain aspects or have a certain length by changing the LLM instruction.

\scriptsize
\begin{verbatim}
    "Bohemian Rhapsody" is a dynamic and genre-blending piece that
   takes listeners on a musical journey. The song begins with a gospel-inspired 
   section, featuring a choir of female vocalists singing harmoniously in a major key, 
   accompanied by a keyboard playing the melody. The tempo is brisk at 171.0
   beats per minute, creating an uplifting and spiritual atmosphere. Around the
   60-second mark, the song transitions into a slower tempo, introducing a male 
   vocalist singing a soft melody accompanied by a piano and a cello playing a single
   note. The song takes on a calming and soothing atmosphere, perfect for a romantic
   movie soundtrack. The chord progression shifts to Bb, A#, Bb, and the beat counts
   to 4, with a tempo of 70.0 beats per minute. At 80 seconds, the song morphs into
   a slow tempo country love song with a female vocalist singing melodically. The 
   song is accompanied by a keyboard playing the melody, a bass guitar playing the
   root notes, and an acoustic guitar playing the melody. The song has a romantic and 
   emotional feel to it, with a chord progression of B, Gm, F#m, B, Gm, F#m, and a 
   tempo of 117.0 beats per minute. At the 100-second mark, the song transitions into
   a bluesy atmosphere with a male vocalist singing in a higher key. The melody
   is played by a harmonica and the rhythm is provided by an acoustic drum. The bass
   plays the root notes of the chords, and the chord progression is Bb, Eb, Bb, Eb,
   Bb, Eb. The tempo remains at 117.0 beats per minute. At 170 seconds, the song takes
   a turn into the realm of instrumental rock, characterized by a medium tempo with
   a groovy electric guitar lead, punchy snare, groovy bass guitar, shimmering hi
   hats, punchy kick and snare hits, and a groovy electric guitar solo. The song
   is in 4/4 time signature and has a chord progression of Cm, Fm, Cm, Fm, Cm, Fm,
   Cm, Fm, Cm. At 200 seconds, the song becomes lively and energetic with a fast
   tempo and a catchy tune. The male vocalists sing in unison, creating a harmonious
   and engaging melody. The song is filled with enthusiasm and passion, making it
   a perfect addition to any party or celebration. The beat is quick and lively,
   with a tempo of 170.0 beats per minute. The song is in the key of A minor. 
   At 260 seconds, the song transitions into a heavy metal/rock song with a fast tempo
   and aggressive electric guitar lead melody. The bass guitar plays the root notes
   of the chords while the acoustic drums provide a simple beat. The atmosphere is
   intense and energetic, perfect for a rock bar. The song is in the key of A# minor
   and has a chord progression of A#m7, Dm7, A#m7, Dm7, A#m7, Dm7, A#m7. At 320
   seconds, the song transitions into a mellow and emotional soft rock song with
   a male vocalist singing melodically. The piano and electric guitar play the main
   melody while the bass guitar and acoustic drum beat provide a simple and slow
   rhythm. The song has a sentimental and mellow atmosphere, making it perfect for
   a romantic movie. The chord progression is Gm, C#m, Gm, C#m, Gm, C#m, Gm, C#m,
   Gm, C#m, Gm, C#m, Gm, C#m, Gm. At 340 seconds, the song transitions into a
   slow tempo instrumental song with a melancholic melody played on the piano. The
   music is emotional and sentimental, creating a poignant atmosphere.  Finally,
   at 360 seconds, the song concludes with a medium tempo instrumental featuring
   a loud crash of timpani and the sound of a marimba in the background. The song
   is in the key of E minor, with a beat of 2 and a chord sequence of E. The tempo
   of this final section is Andante, providing a dramatic and climactic end to this
   musical journey."

\end{verbatim}

\normalsize
We invite the reader to explore to more examples of generated captions on our GitHub page\footnote{\url{https://github.com/AMAAI-Lab/sonicverse}}, where we have also made the model available as open source.

\section{Conclusion}
In this paper, we introduce \model \hspace{-0.5em}, an open-source multi-task music captioning system that integrates caption generation with auxiliary music feature detection to produce richer and more informative descriptions of musical content. By projecting both raw audio and detected features into a shared language space, our model captures not only low-level acoustic attributes but also high-level musical semantics. By training the proposed multitask captioning model on the Magna-Tag-A-Tune dataset \citep{law2009evaluation} with MosiacML \citep{MosaicML2023Introducing} captions and MIRFLEX-based \citep{chopra2024mirflex} features, and finetuning on the MusicBench \citep{melechovsky2023mustango} dataset expanded with MIRFLEX-based features, we can explore the synergy between captioning and feature detection in a unified framework, thus enabling our captioning model to be trained on a smaller, open dataset. Experimental results confirm that this multi-task approach enhances caption quality as well as descriptive depth. The resulting model outperforms state-of-the-art models trained on open data. 

Beyond captioning short audio fragments, our framework facilitates scalable captioning of longer musical works through chaining strategies using a large language model. This opens up promising directions for modeling the temporal structure of music. The resulting \model \hspace{-.5em} is made available online\footnote{\url{https://github.com/AMAAI-Lab/sonicverse}} and as a HuggingFace Spaces demo. 

\begin{ack}
This work has received support from SUTD's Kickstart Initiative under grant number SKI 2021\_04\_06 and MOE under grant number MOE-T2EP20124-0014.
\end{ack}

\bibliographystyle{apalike}   
\bibliography{references}  

\end{document}